\documentclass{PoS}

\usepackage[utf8]{inputenc}
\usepackage{enumitem}

\renewcommand{\deg}{^\circ}
\newcommand{\todo}[1]{}
\renewcommand{\todo}[1]{{\color{red} TODO: {#1}}}

\newcommand{\tohide}[1]{}
\title{Pointing System for the Large Size Telescopes Prototype of the Cherenkov Telescope Array}

\ShortTitle{Pointing System for the Large Size Telescopes Prototype of the Cherenkov Telescope Array}

\author{\speaker{Darko Zarić}${}^, {}^{a,}$ , Stefan Cikota${}^b$, Armand Fiasson${}^c$, Nikola Godinović${}^a$, Koji Noda${}^d$, Robert Wagner${}^e$, Martin Will${}^f$, Quentin Piel${}^c$ for the CTA LST project \footnote{for consortium list see PoS(ICRC2019)1177}\\
	\llap{$^a$} Faculty of Electrical Engineering, Mechanical Engineering and Naval Architecture\\ University of Split, 21000 Split, Croatia\\
	\llap{$^b$} Faculty of Electrical Engineering and Computing, University of Zagreb,  10000 Zagreb, Croatia\\
	\llap{$^c$} Laboratoire d’Annecy de Physique des Particules,\\ Univ. Grenoble Alpes, Univ. Savoie Mont Blanc, CNRS, LAPP, 74000 Annecy, France\\
	\llap{$^d$} Institute for Cosmic Ray Research, the University of Tokyo, 277-8582 Chiba, Japan\\
	\llap{$^e$}	Oskar Klein Centre, Department of Physics, Stockholm University, Albanova
	University Center, SE-10691 Stockholm, Sweden\\
	\llap{$^f$} Max-Planck Institute for Physics, D-80805 Munich, Germany\\
    E-mail: \email{Darko.Zaric@fesb.hr} }

\abstract{The pointing system of the prototype of the Large Size Telescope (LST-1) for the
	Cherenkov Telescope Array observatory, should ensure mapping of the gamma-ray
	image of a point-like source in the Cherenkov camera to the sky coordinates with a
	precision better than 14 arcseconds. Detailed studies of the telescope deformations are
	performed in order to disentangle different deformations and quantify their
	contributions to the miss-pointing, to learn how to correct for them, and finally how to
	design the system for offline and online pointing corrections. The LST-1 pointing
	precision system consist of several devices mounted at the center of the dish:
	Starguider Camera (SG), Camera Displacement Monitor (CDM), two inclinometers,
	four distance meters, and an Optical Axis Reference Laser (OARL), working together
	with the LEDs mounted in a circle around the Cherenkov camera.
	The online pointing corrections are based on a bending model as currently done by
	existing IACTs. The offline corrections will be performed combining measurements
	done by the SG and CDM cameras. SG will provide the position of the Cherenkov
	camera center with respect to the sky coordinates with a precision of 5 arcseconds,
	while CDM will provide the deviation of the telescope optical axis defined by the
	OARL spots with respect to the Cherenkov camera center with a precision better than
	5 arcseconds. Laboratory measurements on dedicated test benches showed that the
	required pointing precision can be achieved for SG, CDM and inclinometer. 
}

\FullConference{36th International Cosmic Ray Conference -ICRC2019-\\
		July 24th - August 1st, 2019\\
		Madison, WI, U.S.A.}

\begin{document}

\section{Introduction}
The prototype of the Large Size Telescope (LST-1) of the Cherenkov Telescope Array (CTA) has been installed in 2018 at the north CTA site located at the Roque de los Muchachos Observatory on La Palma  in the Canary Island and its commissioning should be finished in 2019 \cite{StatusLST_ICRC2019}.

The precision with which a Cherenkov camera coordinate in the focal plane of the CTA telescopes can be mapped to a celestial coordinate is called the telescope pointing precision. 
Knowing the Cherenkov camera position very precisely during favourable observing conditions is needed to achieve the performance requirements of CTA\cite{performance}. Under favourable observing conditions the root-mean-square (RMS) systematic error on the localisation of a point-like source of gamma rays below 100 GeV per axis must be $<10''$ per axis. To achieve this the rms space-angle post-calibration pointing precision for the LST telescopes must be $<14''$.

The LST design is conceived  to provide a large photon collection area sensitive enough to have a low energy threshold of 20 GeV, while the structure should be light as possible to allow a fast reaction to transient events. To achieve these two objectives it is a matter of trade-off, as large collection area needs a big (i.e. heavy) structure, but fast re-positioning requires a light structure which on the other hand leads to deformable structure  affecting the telescope pointing precision. 
%The design of a telescope structure is always a compromise between maximum stiffness and minimum weight(and costs). This holds especially for the LST, as an important physics goal is the observation of transient objects, requiring a fast slewing (and therefore light-weight) telescope structure. Therefore, by design, gravity and wind loads will impose significant deformations on the LST telescope structure.
Due to experiences with the currently working IACTs we already know that we cannot build a perfectly stiff telescope with a reasonable weight. The obtained data must be corrected for all the deformations in the telescope component.
There are several types of deformations that can occur on the telescope such as deformations due to gravity, temperature change, stable wind or by a wind gust.
Currently in the MAGIC telescopes only gravity \cite{MAGIC_pointing}, \cite{Tpoint}  has been considered in the correction methods and the rest of the deformations are corrected for an online correction called "drive bending model" and with an offline precise correction by a Starguider system.

\section{Deformations of the LST}
The	 bending	 model	alone is	 not	 enough	 to	 correct	 for	 the	 deformations	 by	 a	temperature	change, static and		dynamic	winds,	as	these	three	effects	cannot	be	easily	reproduced	in	 advance.
A	gust	or	the	dynamic	component of a	continuous	wind	can change in	a	time	scale	faster	than	a	few	seconds. Since Starguider system will be able to provide corrections with a rate less than a Hz, another system is needed to account for this deformations.
In	order	to	separate	the	camera	displacement	from	the	deflection	in	the	
dish	structure,	Optical Axis Reference Laser (OARL)	spots	will be used as	a	reference,	and	then the	camera	positions	will be measured with	the	help of Reference	LEDs. To correct for the effect by a gust which will have oscillations of $\sim$2 Hz, dedicated Camera Displacement Monitor (CDM) with a rate of 10 Hz will be installed.
This means that OARL will serve as a reference to the dish structure, inclinometers will relate it to the drive system and CDM will relate it to the camera \cite{Pointing_LST}.

It is important to identify all structural deformations contributing to telescope mispointing
and to disentangle them as much as possible using various monitoring techniques.
Apart from global misalignment of the telescope’s azimuth and elevation axes, pointing deviations are expected due to structural deformations of the telescope structure, such as: 
\begin{enumerate}[label={(\arabic*)}]
	\item a rotation or twist of the telescope structure in azimuth direction (not taken account for by the drive	assembly)\label{en:rotation}
	\item a twist of the elevation towers
	\item a twist between the elevation structure (dish and arc) and the elevation motor sitting in the lower \label{en:dish_elevation}
	telescope structure \label{en:dish_drive}
	\item a deformation of the mirror dish \label{en:AMC}
\end{enumerate}

\begin{figure*}
	\center
	\includegraphics[width=0.9\textwidth]{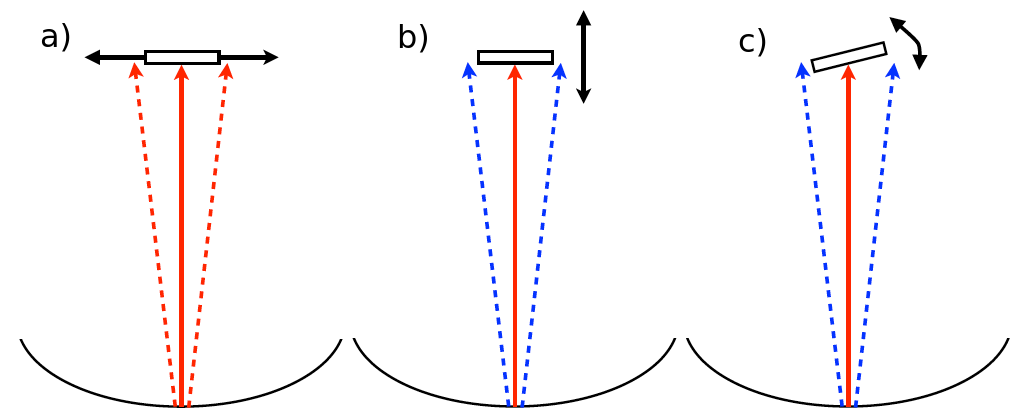}
	\caption{Deformation between the dish and the camera. a) Displacement on the focal surface for x- and 	y-axes (and rotation around z-axis. b) Change of the focal length along z-axis, and c) Tilting of the focal surface, rotations around x- and y-axes}
	\label{fig:dish_camera_deformation}
\end{figure*}

For item \ref{en:rotation} this	deformation	is	expected	to	make	only	a	slow	common	deflection	to	the	dish and	the	arch,	thus	it	can	be	corrected	for	with	the	 Starguider.	
For \ref{en:dish_elevation} inclinometers are installed at the dish center and they will be firmly attached to the OARL and in that way they will directly measure the inclination angle at the dish.
The	deformations \ref{en:AMC} by	the	static	causes	(the	 gravity,	temperature	and	
static	 wind) can	 be	 corrected	 for	 by Active	 Mirror	 Control (AMC).	
Furthermore, deformation of the arc structure may affect the relative orientation of the photomultiplier-tube (PMT) camera w.r.t.the telescope’s optical axis, shown in figure \ref{fig:dish_camera_deformation}, such as:
\begin{enumerate}[label={(\roman*)}]
	\item a vertical or horizontal shift of the focal plane
	\item a rotation of the focal plane around the optical axis
	\item tilts of the focal plane w.r.t. the optical axis
\label{en:tilt}
	\item a shift along the optical axis. \label{en:shift}
\end{enumerate}

% -----------------------------
% *** Deformation between the dish and the camera ***
% -----------------------------

For item \ref{en:tilt} and \ref{en:shift} four distance-meters are	 be installed	 at the	 dish	 centre,	 to	 measure	 the	 distance	 to	 the	 camera	plane,	and	also	its tilting angle.

Finally we	will correct	for	all	the	effects	by	Starguider, which	uses the	sky	field	as	the	reference,	to	achieve the	final	precision	less	than	14	arc-seconds.

Elastic deformations can be well parametrised by a bending model prepared in advance (using special calibration observations where the telescope is pointed at several bright stars homogeneously distributed in the azimuth-elevation plane). Such bending model, however, cannot correct for inelastic deformations of the telescope structure (e.g. due to temperature changes, constant or gusty wind loads). To take these effects into account in the pointing calibration, deformations have to be recorded as much as possible during science data taking. For the LST, installation of the dedicated  monitoring devices is foreseen.

\section{Hardware components}
\subsection{Reference LEDs}
\begin{figure*}
	\center
	\includegraphics[width=0.45\textwidth]{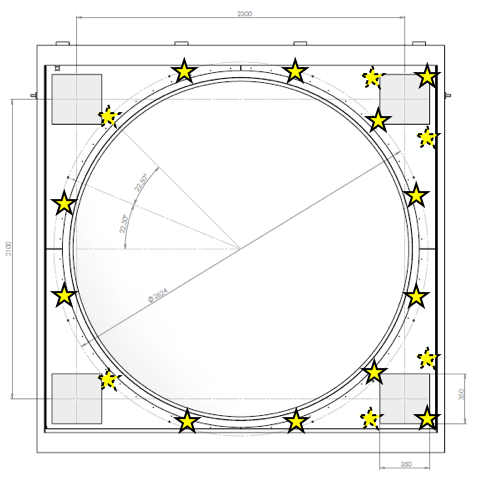}
	\caption{Position of the LEDs on the lid of the PMT camera.}
	\label{fig:LED}
\end{figure*}

Reference LEDS are installed around the PMT camera and are used as a reference, even during data acquisition (DAQ).
There are 12 LEDs mounted in a circle around the PMTs, and 3 more are located in each corner of the PMT camera hood as seen in figure \ref{fig:LED}. 
The LEDs are ROHM SLI-343V8R3F with a dominant wavelength of 630 nm and on each LED there is a cap with a collimated hole with a size of 1 mm.

% -----------------------------
% *** LEDs ***
% -----------------------------

\subsection{Starguider}
This	camera	corrects	for	the	telescope	pointing	with	a	high	precision,	using	known	stars	in	the	catalogues.	It	 sees	 a	 star	 field,	 and	 also	 a	 part	 of	 Reference	 LEDs. Since the Starguider camera will be located in the dish centre,  its line of sight will be inclined by a few degrees w.r.t. the telescope’s optical axis. The	Starguider	can	know	only	the	relative	position	of	 the	 camera	 in	 the	 sky	 field,	 which	 does	 not	 necessarily	 correspond	 to	 the	 pointing	 direction of	 the	telescope	dish.	In	fact,	a	Finite Element Analysis	for	the	LST	structure \cite{FEA}	shows	that,	the	dish	and	the	arch	deform	differently, strongly	dependent	on	the	direction	of	the	wind. For that reason CDM is used. \\
While the LST drive system tracks the source, every second the SG camera  takes the images of stars in its field of view close to the tracking source. 
Images of stars taken by SG camera are analyzed to identify the star and their relative position using the star catalogue to provide mapping of the SG pixels to sky coordinates and the number of correlated stars. Since the SG camera also measures the center of the PMT camera (part of the LEDs mounted on circle around the PMT camera are also in the SG camera field of view)  and the mapping between  the SG camera pixels and the  sky coordinates is known, the sky coordinates of the  PMT camera are also known. 
The Starguider correction will be applied offline.

For the SG a Peltier cooled low-noise camera from Allied vision was chosen. Bigeye G-283 Cool  has resolution of 1928 (H) $\times$ 1452 (V) with pixel size 4.54 microns x 4.54 microns. The Kowa LM50HC lens of 50 mm focal lens gives the field of view (FoV) of 9.938 deg (horizontal) x 7.485 deg (vertical), which results with a resolution of 18.56 arcsec/pix. At 28 m the FoV measures 4.90 m x 3.66 m.
The  software to control SG camera is used over OPC Unified Architecture (OPCUA) server enabling it to take images, control exposure and send images over the Ethernet link to the defined IP address.

\subsubsection{Analysis procedure of the Starguider}
\begin{figure*}
	\center
	\includegraphics[width=0.5\textwidth]{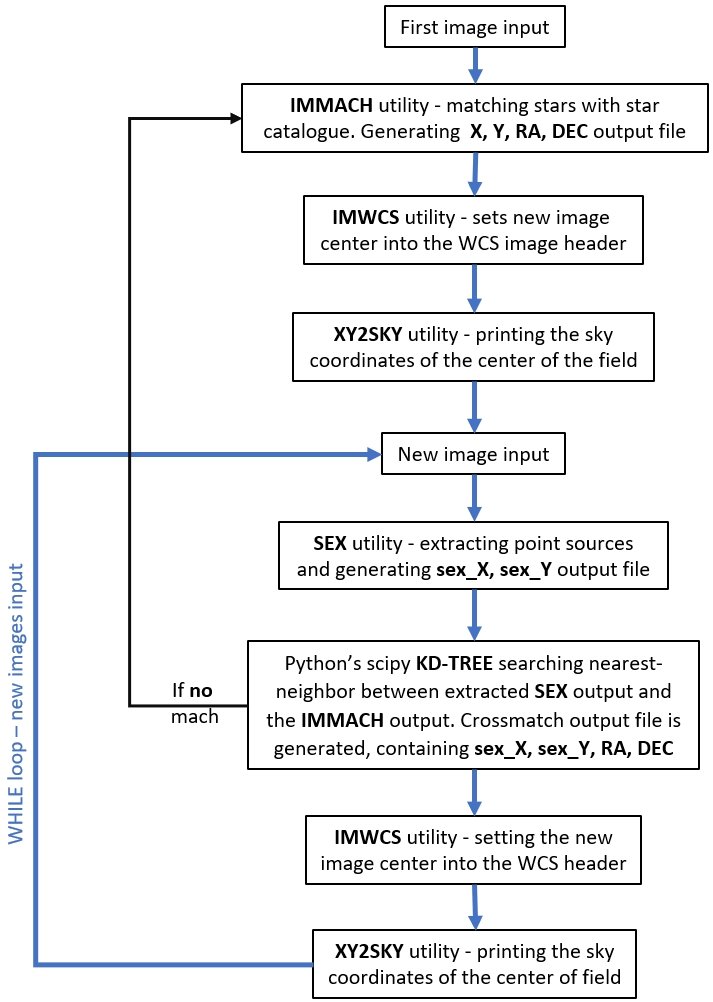}
	\caption{Sketch of the LST starguider software procedure}
	\label{fig:StarguiderAnalysis}
\end{figure*}

The main purpose of the LST Starguider is to find the center of field of the LST. 
The first step towards that is to search the center of field of the star guider camera. The main tools used for image solving and searching the center of field are the WCSTools \cite{WCS}, \cite{SAOimage}.
WCSTools is a package of programs and a library of utility subroutines for setting and using the world coordinate systems (WCS) in the headers of the most common astronomical image format, the FITS format, to relate image pixels to sky coordinates. This software is all written in very portable C.
The WCSTools commands are executed in terminal. Because the LST star guider software consists of a set of different WCSTools procedures, a script in Python was written that executes the WCSTools commands in terminal.
Determining the center of field of an individual image by using the WCSTools is done with the utility "immatch". The “immatch” utility is used for matching stars in a FITS (or IRAF image) with a world coordinate system (WCS) in its header with stars in a reference catalog. The WCS is the relationship between sky coordinates and image pixels and can be described in a standard way in the header of an image. The header of our star guider FITS images was set to be compatible with the standard WCS format while taking (saving) the images. The utility immatch supports different star catalogues: The HST Guide Star Catalog, the USNO UJ Catalog, the USNO A Catalogs, or user-supplied catalogs may be used. 
For software tests with images that were collected at the public observatory Mosor (in the mountains near Split, Croatia), the HST Guide Star Catalog was used. The immatch utility takes between 5 and 8 seconds to reduce the individual images, match the extracted stars with the star catalogue and find the center of field.
Since the main requirement of the LST star guider software is to determine the center of field of individual images within 1 second, the immach utility by itself is too slow, and a different approach was established.
Since the LST telescope is not changing its field of view during the data acquisition of a particular target, 
%and the LST Starguider is mainly designed for precise position monitoring during the LST’s observations, 
%and eventually to perform tracking corrections in order of less than 1 arcminute, 
performing the immatch utility can be avoided for each individual image, and can instead be performed only after every reposition of the LST telescope, while the other star guider images during LST data acquisition will be processed by using SExtractor and Python's SciPy kd-tree tool for quick nearest-neighbor lookup.
%The final procedure used in the LST star guider script is described below step-by-step in figure \ref{fig:StarguiderAnalysis}.
A sketch of the procedure used in the LST star guider script is given in figure \ref{fig:StarguiderAnalysis}.

% -----------------------------
% *** Starguider image analysis ***

\subsection{Optical Axis Reference Laser}
%Optical Axis Reference Laser
OARL defines the optical axis of the LST during science observations. 
Two OARLs are installed in order to provide 2 dimensional information. They are installed at the dish centre, pointing towards screen targets at the edges of the PMT camera. When observed by the CDM, the two reference spots can be compared to the reference LEDs, providing a correction for the PMT camera displacement. 
A stiff unit is built, in which the inclinometer is sandwiched by the two OARLs,
to represent the dish structure. The unit will be installed at the center of the dish with a dummy facet in order to avoid any additional systematic deformation at the central hub.
2 infrared laser chosen as the OARL are made by ProPhotonix. Their wavelength of the emitting light is 780 nm, the power output is 3 mW and they are focusable.

\subsection{Inclinometers}
The inclinometers will be used to measure dish deformations during data taking. Finite element analysis of the LST structure has confirmed that the inclination at the dish center is closely representing that of the whole dish structure. Therefore, it is planned to measure directly the inclination angle at the dish center, by inclinometers firmly attached to OARL.
The   inclinometer   will   be   combined   with   OARL    in   the   hardware   level.
There are two   inclinometers ("Tuff Tilt Digital" by Jewell Instruments),   one   for   the   elevation   angle   (parallel   to   the   optical   axis)    and   one   parallel   to   the   elevation   axis. 
The former one has a wide range of $\pm 50 \deg$ with 14 arsec of precision, while the latter one has more narrow range $\pm 3\deg$, but with a higher precision  of 1 arsec.

\subsection{Camera Displacement Monitor}
Camera Displacement Monitor (CDM) will be located in the central hole of the dish and monitor during science observations the horizontal and vertical displacement of the PMT camera, as well as the rotation of the focal plane w.r.t. 	optical axis defined by the OARL. It will see OARL and Reference LEDs in its FoV. It will use a frame rate of 10 Hz and it will separate the camera displacement from the dish deformation.
The CDM correction will be used offline.

The CDM system is built of: camera, lens, processing unit and a sensor for measuring temperature, pressure and humidity.  Processing unit is Intel NUC  Mini PC NUC5i7RYH running CentOS. 
Camera  is produced by Imaging Development Systems (IDS), and the model is UI-3590CP, which is a CMOS camera connected with USB 3.0 to the processing unit. It can achieve frame rates up to 21.0 fps and its maximal resolution is 4912 x 3684 which amount to 18.10 MPix. The IR filter has been removed from the camera so it can be sensitivity to the OARL spots.
It is equipped with  Edmund Optics 35mm lens. Having focal length of 35mm provides the horizontal FoV of 7 deg and vertical of 9 deg and a sampling of 7 arcsec/pixel.

%The laboratory setup shown in Fig. 1.\todo{Insert this Image?}.
%It was used to test does the CDM could measure a displacement of 0.5 mm from a distance of 20 m, 
%which is the desired precision of 5 arcsecond for CDM.
%\todo{Mention this?}
%It is done in the following way. Take  several thousands images of 12 LEDs on a circle of 2 m diameter  at rate of 10 Hz and analyse them on-fly and plot the distribution of the x and y coordinates of the calculated centre of Cerenkov camera by fitting the measured 12 LEDs coordinates. Then movable table is moved in horizontal direction by 0.5 mm and the same procedure is repeated  giving the distribution of the  x and y Cerenkov camera coordinates see Fig 4.\todo{Insert this?}   The resolving power of CDM to reliably measure the displacement by 0.5 m at 20 m distance (ie. 5 arc seconds)  is defined as difference of the mean of the Cerenkov camera centre  coordinates  distributions divided by the its width. As it is shown in Fig 5.,\todo{Insert this?} the CDM resolving power is almost 10.  

\subsubsection{Bending model}
The same camera will be used for generating the bending models. During the pointing calibration observations it will observe the Reference LEDs on the PMT camera and the image of the bright star projected on a reflective screen in front of the PMT camera.
The only difference is that in this mode the camera will take images with longer exposition times.
It will be controlled using the the ACS (ALMA Common Software) framework and the OPCUA protocol provided by ACTL (Array Control and Data Acquisition System). 

A	star	will be 	tracked	with	the	available	initial	conditions.	Its	direction (Zd,	Az)	is calculated	from	(Ra,	Dec),	and	(Zd,	Az)	is	converted	to	the	motor	steps	with	the	encoders.	The	star	spot	will	be	visible	on	the	central	screen	with	a	shift	with	respect	to	the	optical	axis	(defined	by	the	OARL spots).	The	shift	values	taken	with	all	 the	 tracked	 stars	 in	 a	 large	 range	 of	 (Zd,	 Az)	 are	 stored	 with	 the	 encoder	 and	 inclinometer	measurements.	Then	we	can	generate	a	bending	model,	a	relation	from	the	encoders	 and	 inclinometers	 to	 the	 optical	 axis.	 The correction will be used	 online	 during	 DAQ.	 	 

Parameters to be used in the bending model would be	 as	 follows 1)	 trivial	 offsets	 between	 the	 axes and	 encoders,	 2)	 azimuth	 axis	
misalignment,	3)	non-perpendicularity	of	the	azimuth	and	elevation	axes,	4)	non-perpendicularity	of	the	elevation	and	optical	axis,	5)	small	non-centricity	of	the	axes,	6)	a	zenith-angle	dependent	deflection	of	the	optical	axis.	

%\todo{insert image of the Hood? or insert image of the scheme inside the hood?}

\subsection{Distancemeter}
	
Distancemeters (DMs)   measure   the   distance   and   tilting   of the focal plane w.r.t. the dish centre.   In   general   measurements   for   a   28m    distance   utilize   a   time-of­-flight   sensor   with   a   laser.   
There are 4 DMs (model DL35 made by SICK) that use infrared lasers (laser class 1) in order   not   to   affect   the   DAQ   by   PMTs. They each point in the corner of the PMT camera lid and they each have $\pm 15$mm of accuracy.

\section{Outlook}
%So the   principle of operation is:   OARL   and   the  inclinometers   are   combined   mechanically,   to   be   a   composite   representing   the   dish   structure.   The  inclinometers   connect   the   drive   to   the   optical   system.   CDM   separates   the   camera   displacement  (the   arch   deformation)   from   the   dish   deformation.  
Drawback of two camera solutions, SG camera pointing to the sky  and CDM pointing to the PMT camera, to achieve a  LST pointing accuracy of 14 arc seconds, is the necessity to perform the careful inter-calibration between the cameras. Namely the displacement of the telescope optical axis in respect to the PMT camera centre has to be translated in the sky coordinates of the PMT camera centre measured by SG camera. The transformation matrix between SG camera coordinates and CDM coordinates  will be  determined from the measured coordinates of the same six LEDs  seen by both cameras following the procedure \cite{transformation}. The software tool to inter-calibrate these  two cameras will be tested in the laboratory first by taking image by both cameras of the same 6 LED seen by both cameras without any tilt between them and then with the tilt of 3 degree as it is expected upon mounting on the telescope. Once the housing  with pointing system on the LST dish and proper tilt between two cameras is found out, the calibration procedure tested by  the laboratory measurements  will be done by taking several thousand  images and running on the already developed and tested software for the camera inter-calibration. It is expected that the pointing system will be installed at the LST-1 by the end of August 2019.
%\todo{Mention that all this pointing more complicated than before  IACT}
%\todo{Mention calibration and commisioning still to be done}
\section*{Acknowledgments}
This work was conducted in the context of the CTA LST Project.
We gratefully acknowledge financial support from the agencies and organizations listed here:\\
\verb|http://www.cta-observatory.org/consortium_acknowledgments|\\
and the support by the Croatian Science Foundation (HRZZ) Project IP-2016-06-9782.

%\section{Discussion}

%\begin{Acknowledgements}
%Put acknowledgments here.
%\end{Acknowledgements}

%\begin{thebibliography}{99}
%\bibitem{...}
%....
%
%\end{thebibliography}

\end{document}